\documentclass{aastex631}
\usepackage{amsmath}
\usepackage[figuresright]{rotating}

\newcommand{\angstrom}{\textup{\AA}}

\newcommand{\Fermi}{\emph{Fermi}}

\def \astrid{\textcolor{black}}
\def \alexander{\textcolor{black}}

\def \journalcomments{\textcolor{black}}

\shorttitle{SOL20210717}
\shortauthors{Pesce-Rollins et al.}

\graphicspath{{./}{figures/}}

\begin{document}

\title{The coupling of an EUV coronal wave and ion acceleration in a \Fermi-LAT behind-the-limb solar flare}

\author{Melissa Pesce-Rollins}
\affiliation{Istituto Nazionale di Fisica Nucleare, \\
Sezione di Pisa \\
I-56127 Pisa, Italy}

\author{Nicola Omodei}
\affiliation{W. W. Hansen Experimental Physics Laboratory, \\
Kavli Institute for Particle Astrophysics and Cosmology, \\
Department of Physics and SLAC National Accelerator Laboratory,\\
Stanford University, Stanford, CA 94305, USA}

\author{S\"am Krucker}
\affiliation{University of Applied Sciences and Arts Northwestern Switzerland, \\
CH-5210 Windisch, Switzerland and Space Science Laboratory, University of California, \\
Berkeley, CA 94720-7450, USA}

\author{Niccol\`o Di Lalla}
\affiliation{W. W. Hansen Experimental Physics Laboratory, \\
Kavli Institute for Particle Astrophysics and Cosmology, \\
Stanford University, Stanford, CA 94305, USA}

\author{Wen Wang}
\affiliation{University of Applied Sciences and Arts Northwestern Switzerland, \\
CH-5210 Windisch, Switzerland}
\affiliation{School of Earth and Space Sciences, \\
Peking University, Beijing, 100871, People’s Republic of China}

\author{Andrea F. Battaglia}
\affiliation{University of Applied Sciences and Arts Northwestern Switzerland, \\
CH-5210 Windisch, Switzerland}

\author{Alexander Warmuth}
\affiliation{Leibniz-Institut f\"ur Astrophysik Potsdam (AIP)
An der Sternwarte 16
14482 Potsdam, Germany}

\author{Astrid M. Veronig}
\affiliation{Institute of Physics \& Kanzelh\"ohe Observarory,
University of Graz,
Universit\"atsplatz 5, 8010 Graz, Austria}

\author{Luca Baldini}
\affiliation{Pisa University, Physics Department,
 I-56127 Pisa, Italy}

\begin{abstract}
We present the \Fermi-LAT observations of the behind-the-limb (BTL) flare of July 17, 2021 and the joint detection of this flare by STIX onboard Solar Orbiter. The separation between Earth and the Solar Orbiter was 99.2$^{\circ}$ at 05:00 UT, allowing STIX to have a front view of the flare. The location of the flare was ~S20E140 \journalcomments{in Stonyhurst heliographic coordinates} making this the most distant behind-the-limb flare ever detected in $>$100 MeV gamma-rays. The LAT detection lasted for $\sim$16 minutes, the peak flux was $ 3.6 \pm 0.8 $  (10$^{-5}$) ph cm$^{-2}$ s$^{-1}$ with a significance $>$15$\sigma$. A coronal wave was observed from both STEREO-A and SDO in extreme ultraviolet (EUV) with an onset on the visible disk in coincidence with the LAT onset. A complex type II radio burst was observed by GLOSS also in coincidence with the onset of the LAT emission indicating the presence of a shock wave. We discuss the relation between the time derivative of the EUV wave intensity profile at 193\angstrom\ as observed by STEREO-A and the LAT flux to show that the appearance of the coronal wave at the visible disk and the acceleration of protons as traced by the observed $>$100 MeV gamma-ray emission are coupled. We also report how this coupling is present in the data from 3 other BTL flares detected by \Fermi-LAT suggesting that the protons driving the gamma-ray emission of BTL solar flares and the coronal wave share a common origin.

\end{abstract}


\section{Introduction} \label{sec:intro}

Ever since the first detection of gamma rays from flares whose solar active region (AR) is located behind the visible limb there has been much debate over the acceleration mechanism driving the emission and the location of the emitting region~\citep[See for example][]{1993ApJ...409L..69V,1994ApJ...425L.109B,1999A&A...342..575V,2017FermiBTL}. Two main scenarios have been put forth to explain the observations. In the first one CME-driven shocks are proposed to be responsible for the direct acceleration of the high-energy protons far from the flaring active region, thus allowing the particles to travel from the acceleration site (close to the AR) to the chromosphere on the visible disk ~\citep{cliv93,Plotnikov_2017,Jin2018,Kouloumvakos_2020,2020SoPh..295...18G,wu2021}. In the second one the high-energy protons are accelerated by a flare-related acceleration process, are trapped in large coronal loops~\citep{1994ApJ...425L.109B} ---probably re-accelerated by shock waves~\citep{Grechnev_2018} --- and precipitate slowly to the photosphere~\citep{1991ApJ...368..316R,ryan00,DeNolfo2019}.
Despite numerous works on these rare events a definitive answer to the origin of the high-energy gamma-ray emission associated with the behind-the-limb (BTL) flares is still missing.

\alexander{Globally propagating \astrid{bright fronts} 
observed in the solar corona were first detected by the Extreme ultraviolet Imaging Telescope \citep[EIT][]{1995SoPh..162..291D} by \cite{1997SoPh..175..571M,1997SoPh..175..601D,1998GeoRL..25.2465T}, and were hence named \emph{EIT waves}. They have been subsequently observed by many additional instruments, as well as in other spectral bands, and are thus generally referred to as large-scale coronal waves \citep[for a review, see][]{warmuth2015}. There followed a great debate on the physical nature of the disturbances, the two main competing scenarios being either magnetohydrodynamic (MHD) waves and/or shocks
\astrid{\citep[e.g.,][]{1998GeoRL..25.2465T,warmuth2004b,veronig2008}} or magnetic reconfiguration in the framework of an expanding Coronal Mass Ejection (CME) \citep{delanee1999,chen2002,attrill2007}. The current consensus is that most observational characteristics are consistent with MHD waves or shocks that are launched by the erupting flux ropes that also form the core of coronal mass ejections \citep[cf.][]{long2017,downs2021}.}

\alexander{There is kinematical evidence for different classes of large-scale coronal waves \citep{2011A&A...532A.151W,muhr2014}. Slower waves show a constant speed of roughly 200--300~km\,s$^{-1}$, which is consistent with small-amplitude (i.~e., linear) fast-mode waves propagating at the ambient fast-mode speed in the low corona. In contrast, waves with a large amplitude (often also  characterized by bright and sharply defined wavefronts) start at significantly higher speeds, but show a progressive deceleration down to the speeds of the slower waves. This behavior is consistent with a nonlinear large-amplitude wave or shock where the speed is dependent on the amplitude of the disturbance. Due to the expansion of the wave and the fact that the leading edge travels faster than the trailing one, the amplitude decreases and the wave decelerates, possibly decaying to a linear fast-mode wave.}

\alexander{Fast coronal waves are  highly correlated with metric type II radio bursts \citep{warmuth2004b}. Type~II bursts are characterized by narrow-band emission lanes that drift from high to low frequencies in dynamic radio spectra \citep{wild1950}. They are interpreted as signatures of expanding coronal shock waves at which electrons are accelerated that excite Langmuir turbulence, which is in turn converted to electromagnetic emission \citep[e.~g.][]{mann2018}. In a few events, the notion that at least parts of coronal waves can be shocked and thus able to accelerate particles in the low corona has been proven by radioheliographic observations such as those reported by \astrid{\cite{vrsnak2005}} and \cite{2013NatPh...9..811C}, which show moving radio sources that are co-spatial with the wavefronts. The initial source of the coronal waves and/or shocks is generally identified as the rapid eruption (as well as expansion) of a magnetic flux rope that forms the core of a coronal mass ejection \citep[e.~g.][]{patsourakos2010,veronig2010,kienreich2009,ramesh2012,kumari2017}.} 

Coronal waves have been observed also in coincidence with the behind-the-limb (BTL) gamma-ray solar flares observed by \Fermi-Large Area Telescope (LAT)~\citep{2017FermiBTL} but the connection between the wave and the ion acceleration has never been fully investigated.

In this work we present the observations of the solar flare of July 17, 2021 by \Fermi-LAT and the Spectrometer Telescope for Imaging X-rays \citep[STIX][]{2020A&A...642A..15K} on-board the Solar Orbiter Satellite \citep[SO;][]{2020A&A...642A...1M}. Thanks to the imaging of STIX it was possible to estimate the position of the active region from which this flare originated to be 50$^{\circ}$ behind the eastern visible limb, making this the most distant gamma-ray BTL flare ever observed. This flare was accompanied by a fast CME and a coronal wave was observed by both the
Extreme-Ultraviolet Imaging Telescope \citep[EUVI;][]{wuelser2004} onboard STEREO-A and the Atmospheric Imaging Assembly \citep[AIA;][]{2012SoPh..275...17L} onboard SDO. A complex type II radio burst was also observed by the Gauribidanur Low-Frequency Solar Spectrograph~\citep{ramesh1998,ramesh2011,kishore2014}, which indicates that a shock wave was formed in the corona.

This study presents a clear correlation between the EUV wave characteristics, the type II evolution, and the observed gamma-ray emission. These are compelling evidence that a shock wave could be responsible for the acceleration of protons producing the observed gamma-ray emission for this event.
Furthermore, we find supporting observational evidence indicating that this coupling is also present in the other BTL flares observed by the LAT.  

\section{Observations and Data Analysis}\label{sec:obs}
\subsection{Observational Overview}\label{sec:overview}
On July 17th peaking at 05:07 UTC\footnote{all times related to STIX have been corrected for the light travel time to Earth (+82.5 s)} STIX detected a solar flare with an impulsive phase that lasted about 10 minutes with flare counts seen up to 80 keV. The photon spectral index was rather hard ($\gamma = 2.6\pm0.1$) with a photon flux of $7.1\pm0.1$ ph\,s$^{-1}$\,cm$^{-2}$\,keV$^{-1}$ at 35 keV. These values are in agreement with the correlation reported by \cite{battaglia2005}
indicating that this is a typical flare with a hard spectrum. Below $\sim15$ keV, thermal emission dominates. A single thermal fit to the STIX spectra gives a flare temperature of $17.5\pm0.3$ MK and an emission measure of $28.8\pm2.4\times10^{47}$ cm$^{-3}$ at the non-thermal peak time (05:04:38 UTC). The SO - Earth separation was 99.2$^{\circ}$ at the time of the flare with SO observing behind the east limb. Based on STIX imaging, the flare location as seen from Solar Orbiter is found to be E40S20 thus this flare was located 50$^{\circ}$ behind the eastern visible limb. Hence, the flare is highly occulted with an occultation height radially above the flare of more than half a solar radius. This makes the flare loops completely invisible from Earth. No significant emission was detected from this flare by the \Fermi-Gamma-ray Burst Monitor (GBM)~\citep{Meegan_2009} or by the GOES satellites. Using the time evolution of the emission measure and temperature derived from STIX, we estimate a GOES class of the order of $\sim$M5. The different energy range of STIX and GOES makes this proxy an order of magnitude estimate with the tendency to underestimate the actual GOES class, as STIX is essentially blind to temperatures below $\sim$8 MK. Hence, we can state that this flare was a larger M class flare, but it unlikely reached the X class range. 

The STIX observations were taken during the cruise phase of Solar Orbiter, when all other Solar Orbiter remote sensing instruments were non-operational. Hence, no other context images besides STEREO coverage is available.  The STEREO-A - Earth separation was 45$^{\circ}$ and thus STEREO-A observed this flares as a slightly occulted event. The STEREO-A COR2 observed a CME erupting from the flare location with an estimated linear speed based on COR2 observations of 1500 km s$^{-1}$. Both SDO/AIA and STEREO-A observed a strong coronal wave in EUV. 

The Sun came into the field of view (FoV) of the \Fermi-LAT~\citep{LATPaper} at 05:00 UTC. \Fermi\ detected $>$100 MeV emission starting at 05:15:36 for a duration of $\sim$16 minutes until 05:31:13 UTC. The Sun left the FoV at 05:35 UTC and only an upper limit is found in the last $\sim$4 minutes of the \Fermi-LAT day. A peak LAT flux for this flare of  $ 3.6 \pm 0.8 $  (10$^{-5}$) ph cm$^{-2}$ s$^{-1}$ was found to occur between 05:18:41 and 05:21:08 UTC. No significant emission was detected when the Sun came back into the FoV of the LAT at 06:00 UTC. Such a value for a peak $>$100 MeV flux lies within the average of the distribution of flux values found in the First \Fermi-LAT Solar Flare Catalog \citep{flarecatalog_2021}. When compared to the previously detected \Fermi-LAT BTL flares, the flux value is roughly a factor of ten dimmer than the 2013 October 11 flare (occultation of 20$^{\circ}$) and a factor of 100 dimmer than the 2014 September 1 flare (occultation of 36$^{\circ}$).

\subsection{STIX and STEREO Data Analysis}\label{sec:stix_analysis}

\begin{sidewaysfigure*}[p]
        \centering
        \includegraphics[width=1.0\textwidth,trim=10 180 10 0,clip]{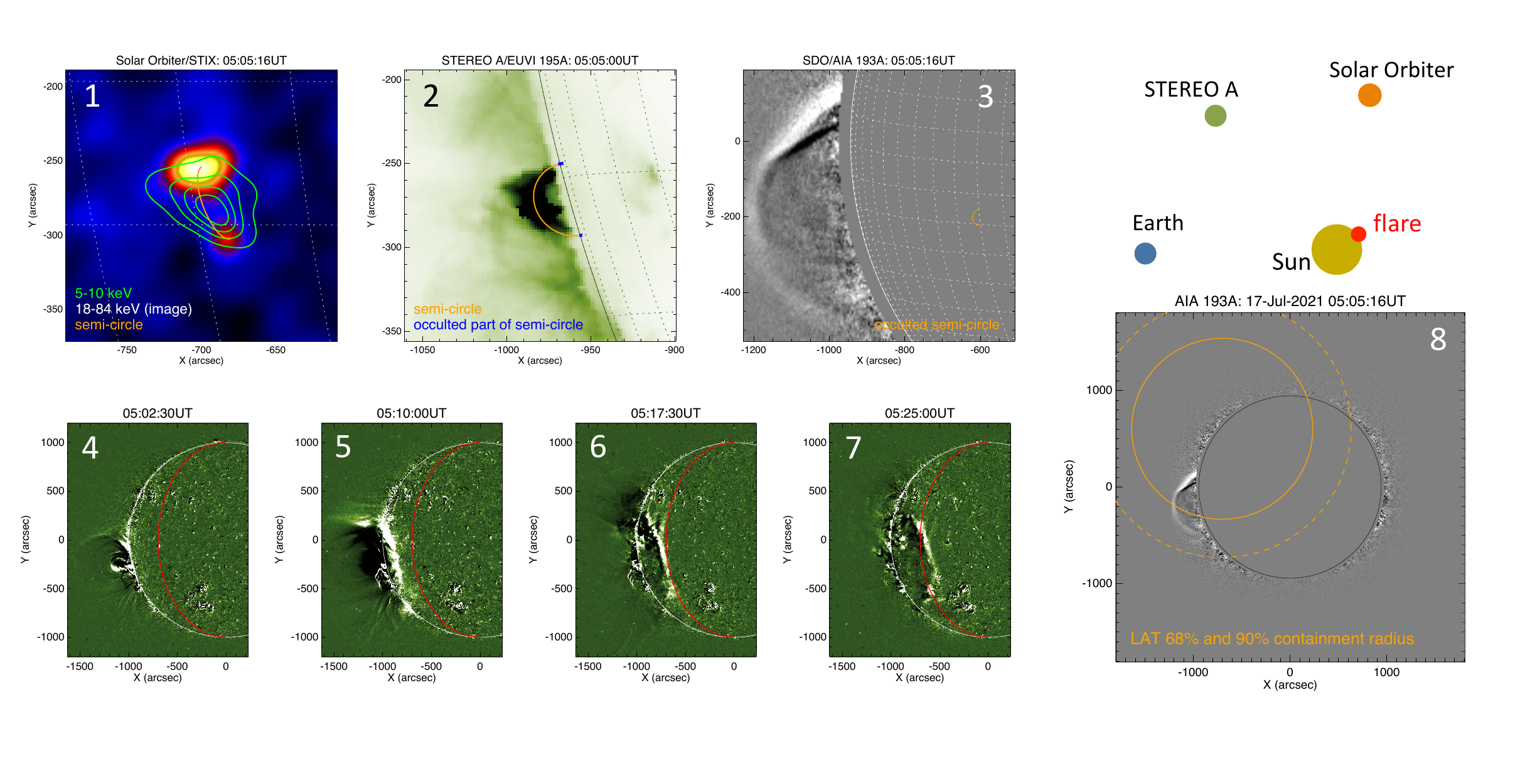}
        \caption{
        Multi-vantage point imaging observations of SOL20210717T05 (the insert on the top right gives the position of Solar Orbiter, STEREO-A, and Earth relative to the flare site): (1+2) The STIX and STEREO-A images reveal the classic picture of a two-ribbon flare. For STIX the flare is seen on disk with nonthermal ribbons connected by a hot thermal loop, while for STEREO-A the flare is at the limb with the flare ribbon being slightly occulted. To guide the eye, a plausible loop (semi-circle standing normal on the surface) has been selected connecting the flare ribbons seen by STIX. The same loop is then shown as seen from STEREO-A.  (3) From the SDO/AIA view, the flare is highly occulted and only the emission associated with the escaping CME is seen in the 193\AA\ running difference image (note that on-disk emission are set to zero for clarity). (4-7) \astrid{EUVI-A 195\AA\ running difference images.}  STEREO-A has a good vantage point to observe the coronal wave as it moves across the disk. The red line marks the solar limb as seen from earth. The coronal wave reaches the limb as seen from earth around 5:17UT. (8) The flare-averaged \Fermi-LAT localization is shown on the same AIA \astrid{193}\angstrom\ running difference as shown in panel 3.
        }
        \label{fig:X_ray_images}
    \end{sidewaysfigure*}

The data and positions relative to the Sun for the telescopes which observed the flare are shown in Figure~\ref{fig:X_ray_images}. The STIX images have been reconstructed with the CLEAN~\citep{1974A&AS...15..417H} algorithm using a clean beam size of 10 arcsec. The finest two subcollimators which are not yet fully calibrated have been excluded. The STIX data used in this paper has been taken during the Solar Orbiter’s cruise phase outside the nominal science windows when Solar Orbiter was at 0.85 AU. The STIX aspect system is designed to provide absolute pointing only at distances closer than 0.75 AU. Hence, we do not have the full accuracy to place STIX images to arcsec precision for this event, but the absolute location accuracy is nevertheless known to about ~2 arcmin. For Figure~\ref{fig:X_ray_images}, we used the flare loop location as seen by STEREO-A for co-alignment by eye. A shift of the STIX thermal image to the north by 62 arcsecs and in west by +20 arcsecs gave a good agreement, see Figure~\ref{fig:X_ray_images}. In any case, for the conclusion drawn in this work, this correction is not critical. In the nonthermal (18-84 keV) image two distinct footpoints are evident, with the northern more pronounced. The thermal emission seen in the 5-10 keV image connects the two footpoints consistent with a classic two ribbon flare geometry. At the same time, the STEREO-A - Earth separation was 45$^{\circ}$ and the second panel in Fig.~\ref{fig:X_ray_images} shows an image taken by its EUVI telescope at 195\AA. From this perspective, the flare loop is clearly visible above the limb although the footpoints are slightly behind the optical disk (the occulted part of the reference loop is shown in blue). 
The last panel in the first row of Fig.~\ref{fig:X_ray_images} shows the CME observed by SDO/AIA in the same wavelength band and almost at the same time. The flare is entirely behind the limb as observed by Earth, nevertheless, a strong coronal wave is associated with this event. The second row of Fig.~\ref{fig:X_ray_images} shows four distinct running difference maps at different times showing the progression of the coronal wave as observed by STEREO-A. The red arc indicates the limb of the Sun as seen by Earth. The coronal wave crosses the limb at approximately 05:17 UT (see last two images)\footnote{A movie of the Sun at 211\AA\ as recorded by SDO-AIA can be found at: \url{https://tinyurl.com/y6jdv54w}}.

To study the time evolution of the coronal wave, we estimate its intensity by integrating the STEREO-A 193\angstrom\ images summed over the area visible from Earth. The area has been further restricted to only include pixels where the wave is clearly visible, i.e. high-latitude regions have been excluded. The same analysis was done for SDO/AIA, however, the analysis is less clear due to the strong projection effects near the limb. The STEREO-A observations therefore provide a much cleaner measurement despite the lower temporal resolution. 
Assuming the wave originates from the flare site and travels along the solar surface, we can define a sector encircling the wave front and the origin of the sector is the flare site. For each running-difference frame, we sum up intensities of pixels which have the same distance to the origin, therefore, we have an intensity-distance profile for the wave front \citep[e.g.][]{vrsnak2002,warmuth2004b}
. We perform a Gaussian function fitting to the intensity-distance profile and obtain the location of the wave front and the corresponding 1-$\sigma$ estimation for each running-difference frame. 
Finally, we summarize all the frames and obtain the distance-time
information representing the propagation of the wave front. We perform a linear fit to the locations and times of the wave front and derive its propagation speed. The estimated speed of the wave front for the 2021 July 17 flare is found to be 269$\pm$12 km s$^{-1}$.

\subsection{LAT Data Analysis}\label{sec:lat_analysis}

We performed an unbinned likelihood analysis of the \Fermi-LAT data within the Multi-Mission Maximum Likelihood (\texttt{3ML})\footnote{\url{https://threeml.readthedocs.io/en/stable/index.html}} framework using \texttt{fermitools}\footnote{\url{https://github.com/fermi-lat/Fermitools-conda/wiki}} version 2.0.8. We selected Pass 8 Source class events from a 10$^{\circ}$ circular region centered on the Sun and within 100$^{\circ}$ from the local zenith (to reduce contamination from the Earth limb). 

First we performed a time integrated analysis from 05:00 to 5:35 UT, optimizing the location of the gamma-ray emission by applying \texttt{gtfindsrc} selecting events above 100 MeV. The best location was at R.A. = 116.720, Dec = 21.356 (degrees, J2000), with 68\% (90\%) containment radius (i.e. statistical localization errors) of 0.26 (0.37) degrees. This location corresponds to helioprojective longitude = -701'' and latitude 603'' (at 2021-07-17UT05:25:09) and is located at 0.27 degrees from the center of the Sun. The centroid of the gamma-ray emission together with the 68\% and 90\% containment radii are shown overlayed on the SDO/AIA running difference image of the Sun in Figure~\ref{fig:X_ray_images}. The uncertainty of the localization of the LAT emission is rather large due to the low photon count statistics, nevertheless, the source location clearly favours the eastern limb, consistent with the associated flare location.

With the best localization as center of the region of interest (ROI), we fitted three models to the \Fermi-LAT gamma-ray data selecting all the events above 60 MeV to better constrain the shape of the spectrum at low energies. The first two, a pure power law (PL) and a power-law with an exponential cut-off (PLEXP)\footnote{The definition of the models used can be found here: \url{https://fermi.gsfc.nasa.gov/ssc/data/analysis/scitools/source_models.html}} are phenomenological functions that may describe Bremsstrahlung emission from relativistic electrons. The third model uses templates based on a detailed study of the gamma rays produced from decay of pions originating from accelerated protons with an isotropic pitch angle distribution in a thick-target  model \citep[updated from][]{murp87}. In all the three analyses, the background is modeled by a fixed contribution coming from the galactic gamma-ray emission (described by the standard template available in the \texttt{fermitools}), and by an isotropic emission describing the unresolved particle background (also described by the standard available template). This latter background component is left free to vary as it encompasses the background variation due to orbital modulation.

We rely on the likelihood ratio test and the associated test statistic TS \citep{Mattox:96} to estimate the significance of the detection. The TS of the power-law fit (TS$_{\rm PL}$) indicates the significance of the source detection under the assumption of a PL spectral shape and the $\Delta$TS=TS$_{\rm ALT}$ - TS$_{\rm PL}$ quantifies how much an alternative model improves the fit. Note that the significance in $\sigma$ can be roughly approximated as $\sqrt{\rm TS}$.

\begin{figure*}[ht!]   
\begin{center}
\includegraphics[width=0.45\textwidth,trim=0 0 0 0,clip]{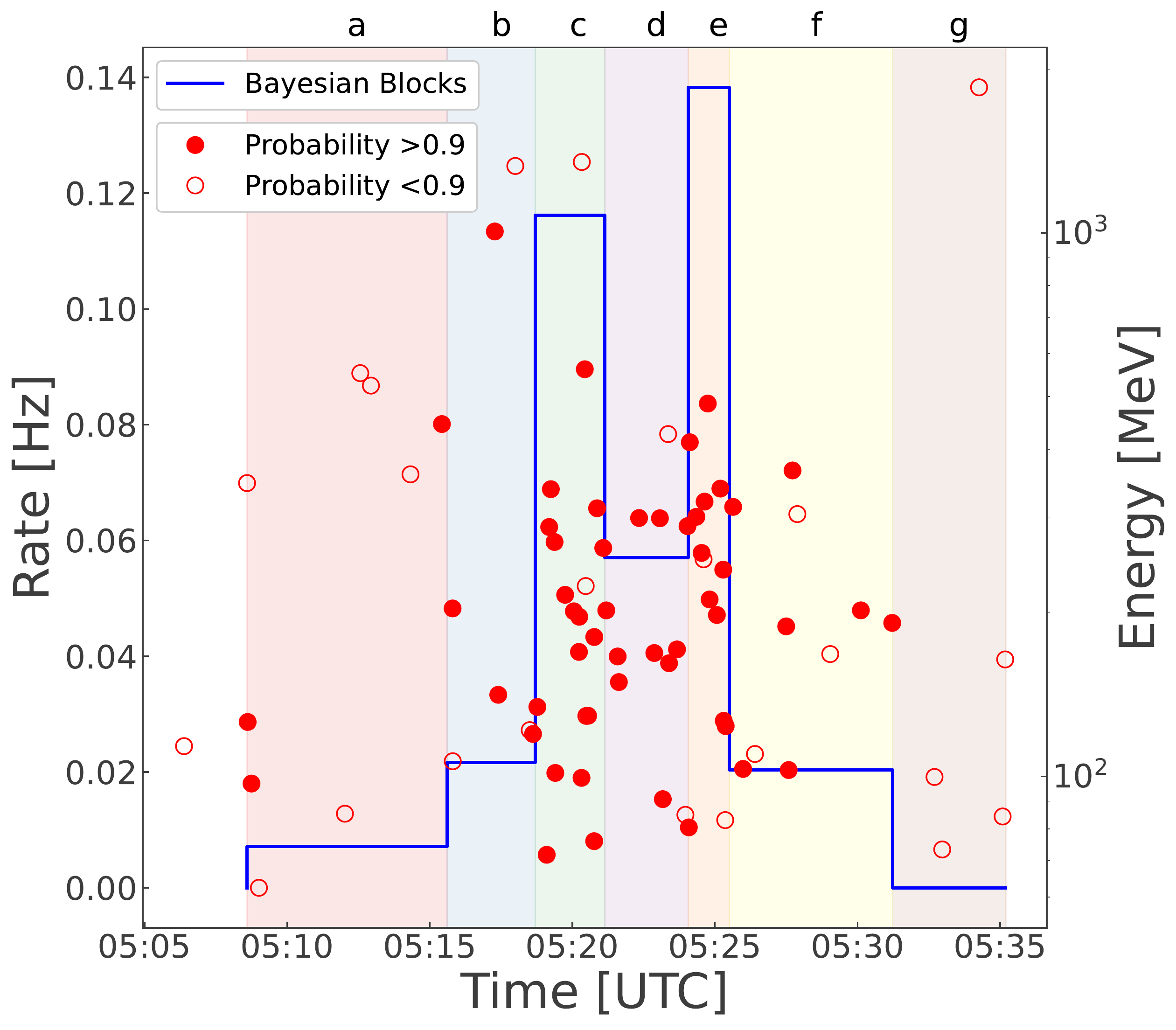}
\includegraphics[width=0.4\textwidth,trim=0 0 0 0,clip]{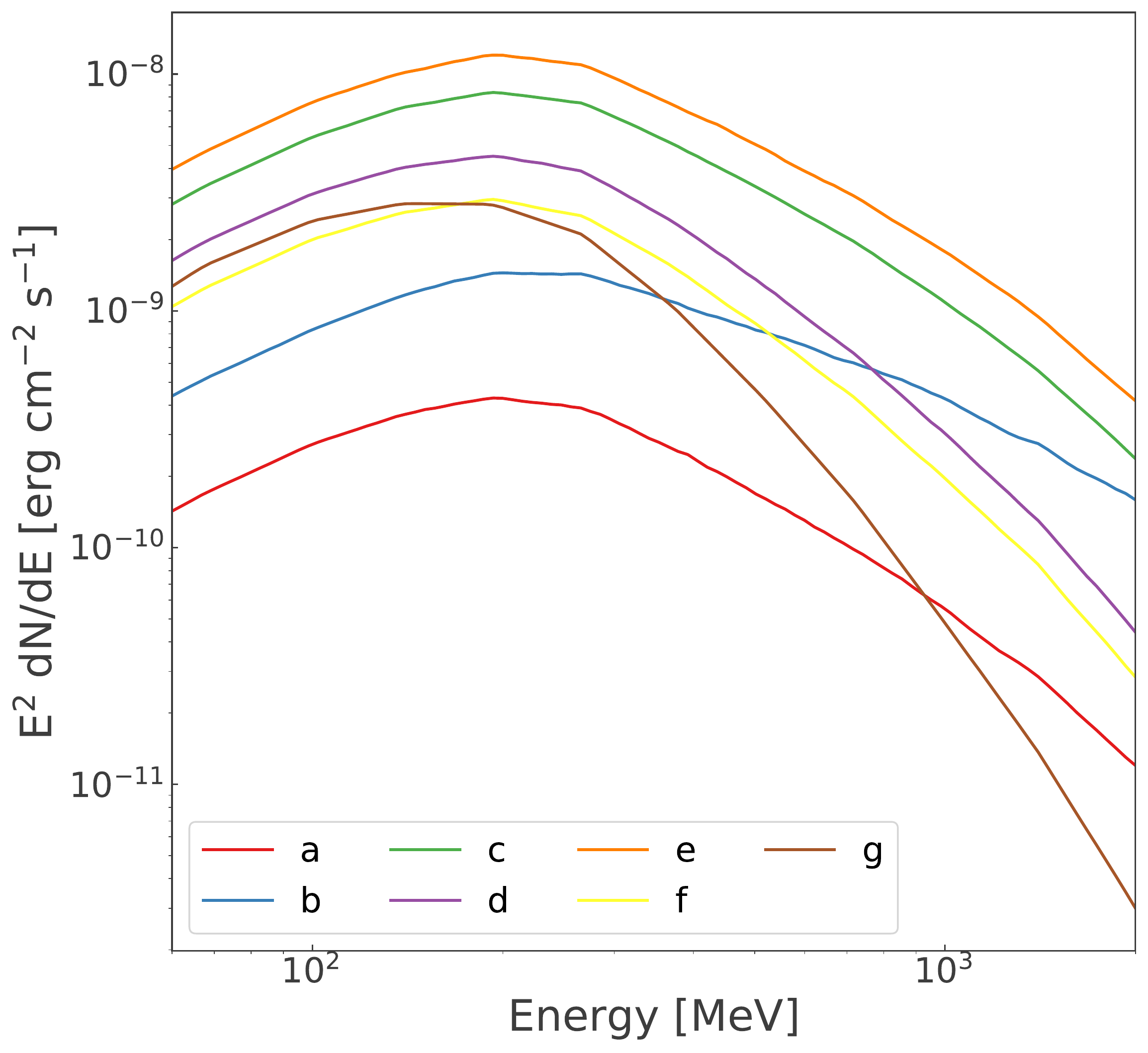}
\caption{\Fermi-LAT time resolved analysis of the flare. Left: Circles are the arrival time and the energies of all the events in the ROI with filled circles highlighting the ones with probability $>0.9$ to be associated with the Sun. Photons represented by the red filled circles were binned using the Bayesian Blocks algorithms and their rate in each time bin are shown with the blue line. The letters annotated in the top of the figure identify the time interval and the unbinned likelihood results for each interval are reported in Table~\ref{tab:spectral_results}.  Right:\Fermi-LAT spectra obtained in each bin. The color of each spectra corresponds to the shaded interval in the left panel.}
\label{fig:events_spectra} 
\end{center}   
\end{figure*}

For the power-law model, we obtained TS$_{\rm PL}$=273, while for the exponential cut-off  we obtained an improvement of $\Delta$TS$\approx$16, suggesting that the curved model is preferred with $\sim$4 $\sigma$ significance.
We used the results from the time integrated likelihood analysis using the exponential cut-off model to compute the probability that each event is associated with the Sun (using \texttt{gtsrcprob} tool available in the \texttt{fermitools}). In the left panel of Fig.~\ref{fig:events_spectra} we show all the events in the 10-degree ROI highlighting the ones with high probability (p$>$0.9) to be associated with the source. 
To define the binning for the time-resolved spectral analysis we applied the Bayesian Blocks algorithm~\citep{scargle_2013}. In order to only incorporate statistical fluctuations from the events that are associated to the flare, we restrict our analysis to events with high probability to be associated to the source. We obtained 7 bins (from \emph{a} through \emph{g}) and we performed a spectral analysis in each bin, using the physically motivated pion model that provides direct information on the evolution of the underlying proton distribution.

\begin{figure*}[ht!]   
  \begin{center}                             
    \includegraphics[width=0.9\textwidth,trim=0 0 0 0,clip]{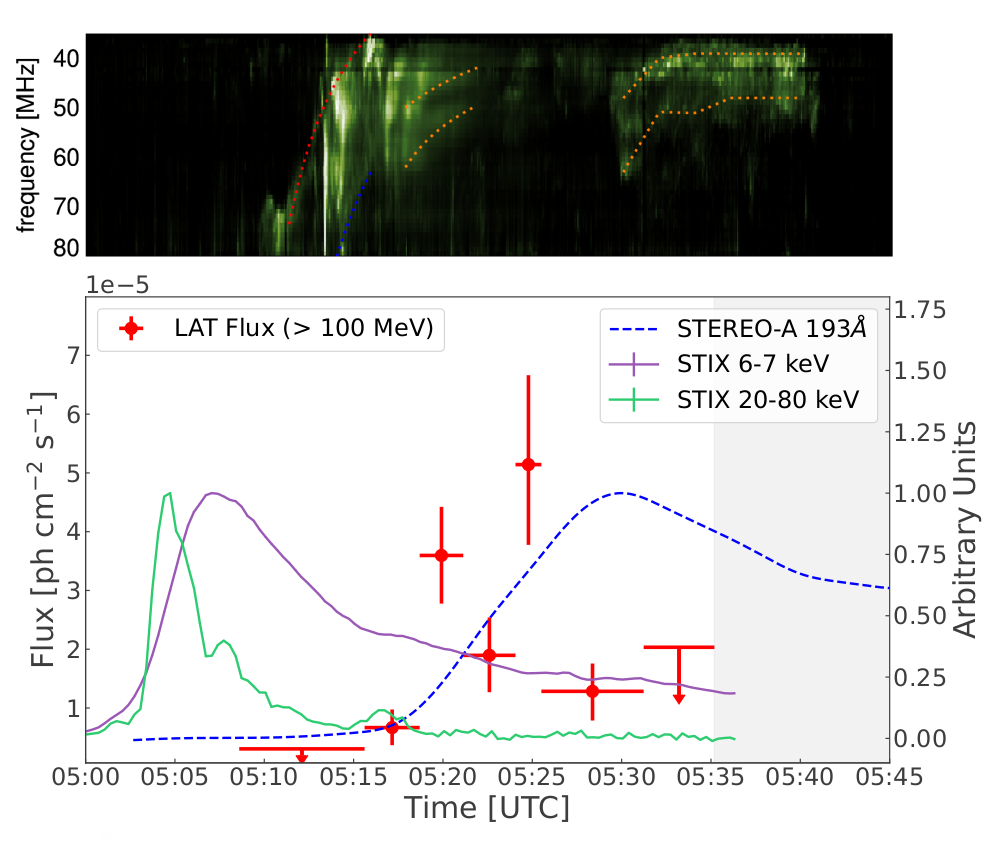}
\caption{Multiwavelength light curves of SOL20210717. Top panel: dynamic radiospectrum from the Gauribidanur Low-Frequency Solar Spectrograph showing a complex type II radio burst. Bottom panel:  \Fermi-LAT $>$100 MeV flux points (red markers), normalized STIX time profile in 6-7 keV (purple solid line) and 20-80 keV (green solid line) energies and STEREO-A 193\angstrom\ of the coronal wave (blue dashed line) as seen from visible disk. The gray band represents the time in which the Sun left the \Fermi-LAT FoV. \journalcomments{The Sun came into the \Fermi-LAT FoV at 05:00 UTC.}}
\label{fig:flux_lc} 
\end{center}   
\end{figure*}

In Tab.~\ref{tab:spectral_results} we summarize the results for the time resolved likelihood analysis. If the significance of the source TS$_{sun}$ is below 9 ($\approx3\sigma$), we compute the value of the flux upper limit. This happened in the first and last bins: in particular, in the first bin we only had a marginal detection, with TS$_{sun}$=8.8. The flux then rapidly increased showing a double-peak structure. In the last bin, even if the Sun was within the FOV of the LAT, it was not detected. In the right panel of Fig.\ref{fig:events_spectra} we show the evolution of the energy spectra. The time profiles of this flare are shown in Figure~\ref{fig:flux_lc} where the \Fermi-LAT flux points obtained in the time resolved analysis are displayed together with the normalized light curves from STIX (6-7 keV and 20-80 keV) and the EUV wave intensity profile at 193\AA\ as observed by STEREO-A.

\begin{table}[h!]
    \centering
    \begin{tabular}{c|l r r r}
    Interval & Time window & Flux$_{0.1-1.0\rm GeV}$ & proton index & TS$_{sun}$ \\
             &  &  cm$^{-2}$ s$^{-1}$   &   &  \\
    \hline
    \hline
a & 05:08:35 - 05:15:36 & $<$3.0$\times10^{-6}$         & $<$4.0  & 8.8 \\
b & 05:15:36 - 05:18:41 & (7$\pm$ 3)$\times10^{-6}$     & 3.3 $\pm$  0.9 & 20.1 \\
c & 05:18:41 - 05:21:08 & (3.6$\pm$0.8)$\times10^{-5}$  & 4.0 $\pm$  1.0 & 121.9 \\
d & 05:21:08 - 05:24:03 & (1.9$\pm$ 0.6$)\times10^{-5}$ & 4.7 $\pm$  1.1 & 59.6 \\
e & 05:24:03 - 05:25:30 & (5$\pm$1)$\times10^{-5}$      & 4.0 $\pm$  1.0 & 100.0 \\
f & 05:25:30 - 05:31:13 & (1.3$\pm$ 0.5)$\times10^{-5}$ & 4.7 $\pm$  1.0 & 36.3 \\
g & 05:31:13 - 05:35:11 & $<$2$\times10^{-5}$           & $<$6.0         & 3.4 \\
    \end{tabular}
      \caption{Results for the time resolved likelihood analysis. In each interval, we report the time window for the analysis, the photon flux between 100~MeV and 1~GeV (in cm$^{-2}$ s$^{-1}$), the proton spectral index, and the significance of the source. If the significance is $<$9, we report the value of the flux upper limit.}
    \label{tab:spectral_results}
\end{table}

\section{Relation between coronal wave and LAT flux}\label{sec:relation}

The high-energy emission detected by the \Fermi-LAT starts at approximately 05:15 UT, 10 minutes after the peak of the impulsive emission detected by STIX. If there was any high-energy emission associated with the impulsive phase, this would have been occulted and therefore not visible (as is confirmed in Figure~\ref{fig:X_ray_images}).
This onset roughly coincides with the time when the coronal wave crosses the limb and becomes visible from Earth, at approximately 5:17 UT as estimated from SDO-AIA and STEREO-A images.

As can be seen in Figure~\ref{fig:flux_lc}, the LAT light curve shows hints of a double peak structure. To further investigate this  we fit the LAT flux points testing two different models using \texttt{3ML} and adopting \texttt{multinest} as Bayesian inference tool \citep{Multinest1,Multinest2,Multinest3}.  
The single peak model is defined as:
\begin{equation}
\rm P(t\,|\,K,t_{p},\tau)=  K \left(\frac{t}{t_{p}}\right)^{t_{p} / \tau} \exp\left(-\frac{t - t_{p}}{\tau}\right)
\label{peak1}
\end{equation}
while the double peak model is simply the sum of two peak functions: $\rm P(t\,|\,K_1,t_{p1},\tau_1,K_2,t_{p2},\tau_2)= P(t\,|\,K_1,t_{p_1},\tau_1)+P(t\,|\,K_2,t_{p_2},\tau_2)$.

We use uniform priors for every parameter, in particular for  $\rm K,K_{1},K_{2}$: uniform from 0 to 1$\times10^{-5}$ cm$^{-2}$ s$^{-1}$, $\rm t_{p}, t_{p1},t_{p2}$ : uniform from 5:20 to 5:30 UTC, and $\rm \tau, \tau_1, \tau_2$: uniform from 0 to 100 seconds.
Using the simplest model we obtain that the peak of the high energy emission happens at $\rm t_p$=5:23:20 ($\pm$ 1 minute) and with a time scale $\tau$=66$\pm$20 s.  Using the more complex model, the positions of the two peaks are instead at $\rm t_{p1}$ = 05:22:40 ($\pm$ 3 minutes) and $\rm t_{p2}$ = 05:23:30 ($\pm$ 2 minutes) with  $\tau_1 = 7\pm6 s$ and $\tau_2 = 56\pm4 s$. Using the Akaike information criterion \citep[AIC][]{AIC} the second model is preferred by a $\Delta$AIC=26, nevertheless, using the Bayesian information criterion \citep[BIC][]{BIC} that penalizes more complex models (with more parameters), we only obtain a $\Delta$BIC=2. Similar results are obtained comparing the logarithms of the Bayesian evidence provided by \texttt{multinest}\footnote{Respectively log(Z) = $-$4.625 for the one-peak and log(Z) = $-$4.583 for the two-peaks model.}
These values suggest that we cannot favor the second more complex model and therefore we conclude that the single peak model (Equation~\ref{peak1}) offers a reasonable description of the high-energy flux. 

To estimate the systematic uncertainty in the STEREO-A profile caused by unrelated fluctuations in the observations, we compute the standard deviation of the points before the onset of the coronal wave reaches the visible disk (before  5:15 UT). We then randomly add this systematic error (estimated to be 0.016 a.u\journalcomments{\footnote{The same arbitrary units used in Figure~\ref{fig:flux_lc}}}.) to the intensity points of the normalized STEREO-A light curve (the statistical uncertainties are negligible). We then describe each of the time profiles obtained in this way using a piecewise polynomial function (spline\footnote{The degree of smoothing of the spline is equal to 3.}) and we compute its derivative. We compute the 16$^{\rm th}$ and the 84$^{\rm th}$ percentiles obtaining the 68\% contour band.

To investigate whether the coronal wave and the gamma-ray emission share a common origin we searched for additional similarities between these two phenomena. The gamma-rays observed by the LAT are produced by ions interacting with the dense photosphere. Under the assumption that the disturbance responsible for accelerating the protons is the same that is driving the coronal wave, then the time derivative of the intensity enhancement  of the coronal wave should provide information on the rate of particles being accelerated and in turn precipitating to the surface of the Sun.

\begin{figure*}[ht!]   
\begin{center}
\includegraphics[width=0.45\textwidth,trim=0 0 0 0,clip]{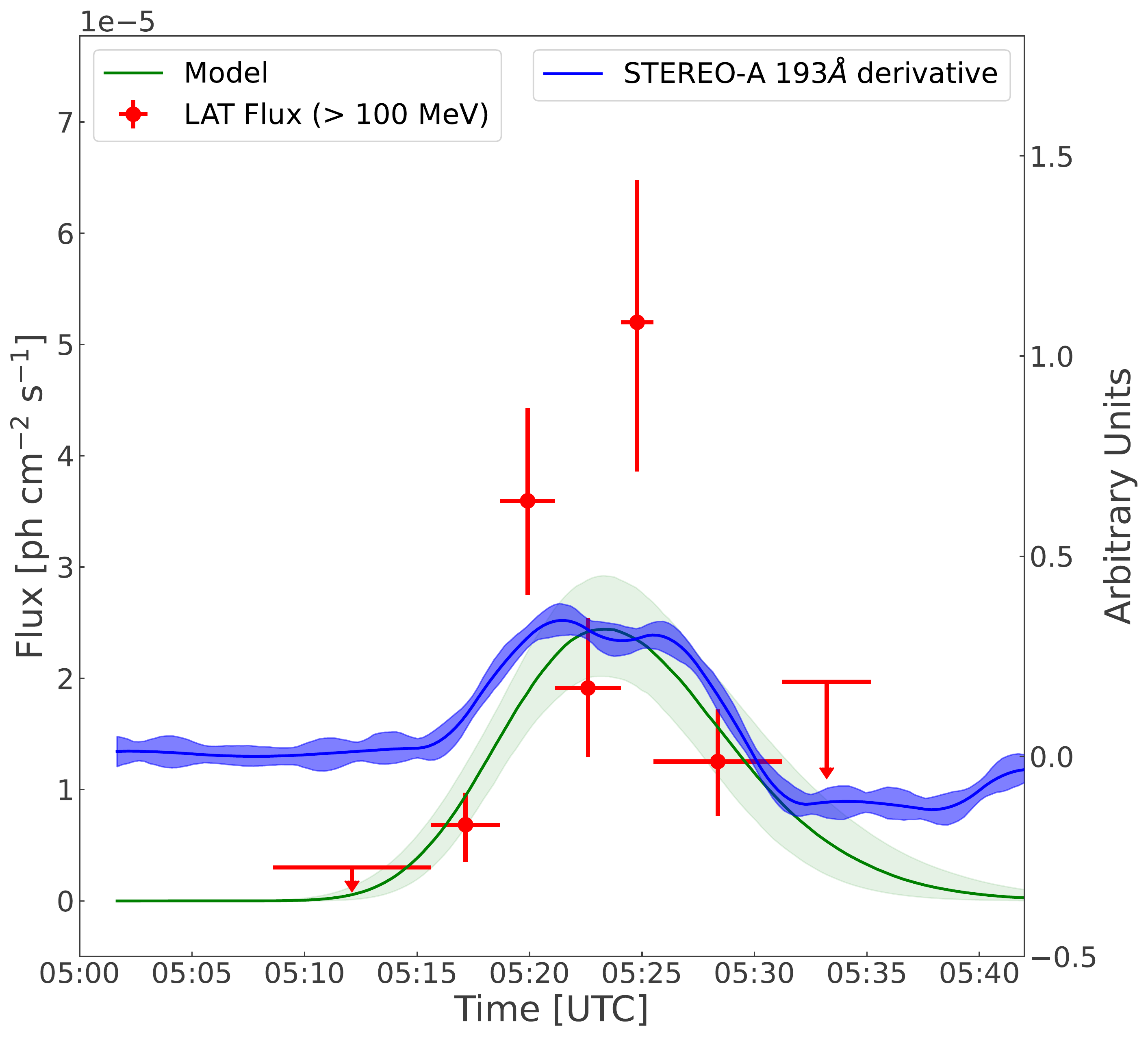}
\includegraphics[width=0.45\textwidth,trim=0 0 0 0,clip]{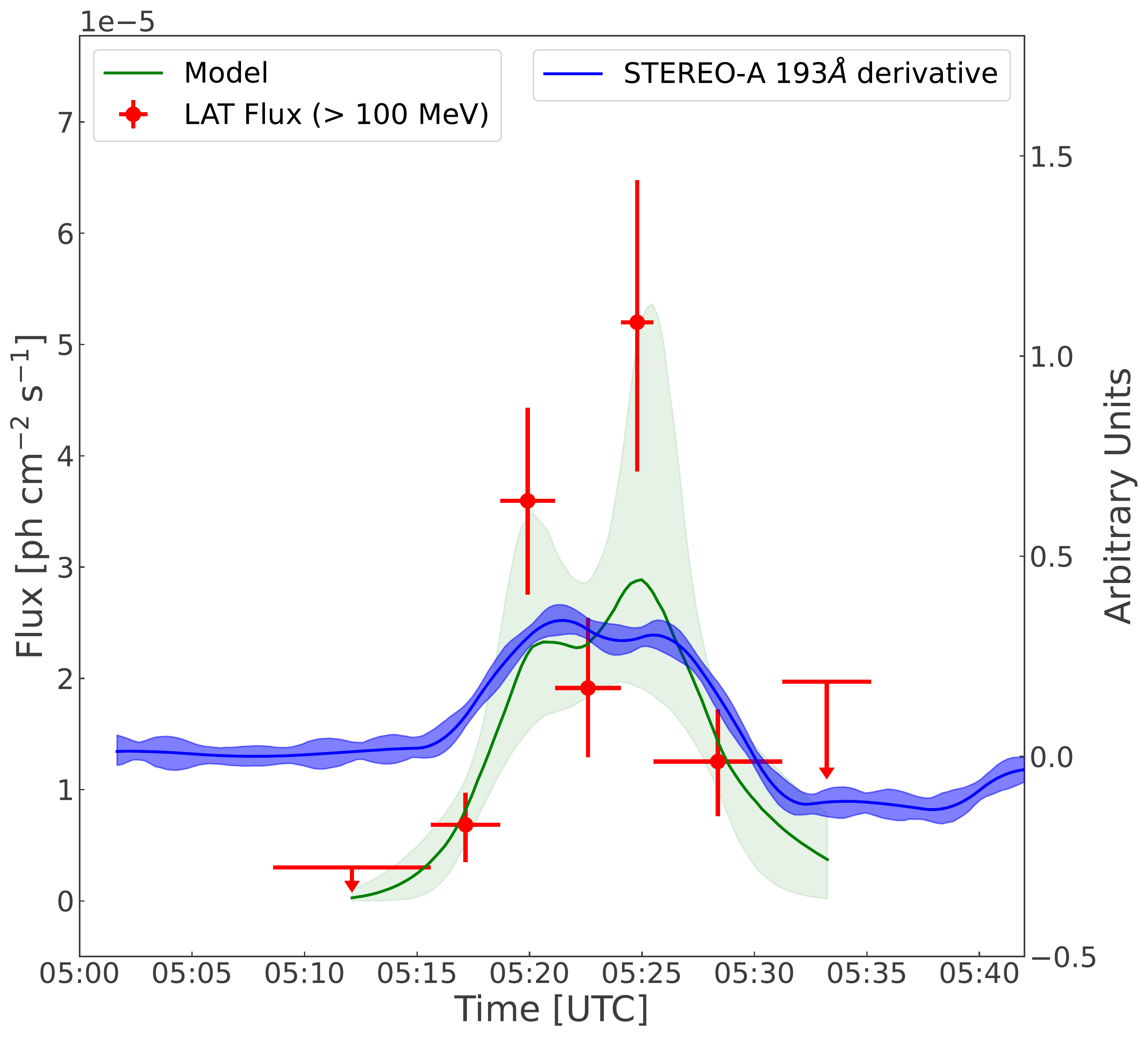}
\caption{Flux $>$ 100 MeV detected by the LAT is shown as red data points. 
The results of the fit with two different models (left: single peak, right: double peak) are shown in green with green contours indicating the 68\% confidence level. Based on the Bayesian information criterion, the single peak model (left panel) is preferred over the double peak (right model). The blue curve is the derivative of the STEREO-A 193\AA\ time profile with the 68\% containment band. The derivative of the STEREO-A light curve peaks at 05:21 UT and the LAT flux is found to peak at 05:23 $\pm$ 1 minute (from the single peak model).}

\label{fig:lcfit} 
\end{center}   
\end{figure*}

The results of the fit to the $>$100 MeV flux points with a single peak (double peak) are shown in the left (right) panel of Fig.~\ref{fig:lcfit}, where the green bands correspond to the 68\% confidence level. The time derivative of the STEREO-A light curve and its 68\% contour band is overlaid in blue in each panel. The position of the LAT flux peak, $\rm t_p$, is nicely aligned with the maximum of the derivative of the STEREO-A light curve which is found to occur at 05:21 UT.

It is interesting to note that the time derivative of the STEREO-A light curve appears to have a double-peak structure coincident with the position of the peaks of the double-peak model $\rm t_{p1}$ and $\rm t_{p2}$ (i.e. the LAT flux points in the c) and e) intervals). 
In order to test whether the systematic uncertainties in STEREO-A could induce such a feature we performed a simulation of a toy model null hypothesis. The toy model light curve is represented by a sigmoid whose derivative is a normal distribution. We introduce the estimated systematic uncertainties to our toy model and performed 10k realizations of the derivative and searched for double peak structures in the resulting distribution. We find that in only 2\% of the cases the derivative is not a smooth bell shape thus suggesting that the structures observed in the STEREO-A derivative could be real and not due to systematic uncertainties, however given that the single peak model cannot be ruled out statistically we cannot derive any additional conclusions on the presence of the structures.
 
 While the striking temporal correlation between the LAT flux and the coronal wave supports the notion of proton acceleration at the wave, we have to note that the moderate speed of the wave does not place it into the kinematical class that is associated with large-amplitude disturbances of shocks \citep{2011A&A...532A.151W}. However, while we have measured the wave speed in the low corona, the disturbance can be faster at larger heights \citep[ e.~g.][]{veronig2010}, where wave signatures are more difficult to observe due to the exponential density fall-off \citep[see e.~g.][]{wu2021}. We have therefore checked the dynamic radiospectra for the appearance of type II radio bursts, which are signatures of shock waves in the corona. The top panel in Fig.~\ref{fig:flux_lc} shows a dynamic radio spectrum provided by the Gauribidanur Low-Frequency Solar Spectrograph. This reveals that our event is associated with a complex type II-like radio burst. In contrast to a textbook-like type II burst that shows a clearly defined emission lane (which may be additionally split into two sub-bands), this burst shows several patchy narrow-band emission features. These are seen from 05:10 to 05:20~UT and have drift rates that are typical for type II bursts. The main feature is indicated by a red dotted line. There is a closely corresponding emission lane (blue line) at 1.8 times the frequency, which would still be compatible with second harmonic emission. If we accept this, we can derive a speed of $\approx$1,000~km\,s$^{-1}$ (against the density gradient) for the radio burst source. From 05:18 to 05:22~UT, an emission lane with a split band that shows a considerable slower drift is observed (orange lines). Finally, another split band is seen from 05:30 to 05:40 UT (orange lines as well). The latter feature does not show any drift after its early phase. We interpret these two bursts as being generated by the flanks of the expanding shock. Propagation at a constant height will lead to a non-drifting type II emission~\citep{2020A&A...635A..62M,Chrysaphi_2020}. This event is very similar to the one discussed by \cite{2013NatPh...9..811C}, where radioheliographic observations were available that proved the lateral propagation of the source.
 
 Thus, the radio observations prove that the coronal disturbance was at least partly shocked during the event, although the wave signatures observed in the low corona were most probably not shocked. The band-split observed in the event can be  interpreted in terms of emission coming from up- and downstream of the shock \citep[e.~g.][]{vrsnak2001}. The band-splits seen in this event correspond to density jumps of 1.4-1.7, quite typical for type II bursts. This corresponds to Alfvenic Mach numbers of a similar magnitude.
 
 \subsection{Relation with other BTL LAT flares}
 \label{sec:otherbtls}
Given that the LAT flux peaks in coincidence with the time in which the derivative of the coronal wave peaks, 
suggesting that the protons responsible for the gamma-ray emission could be accelerated by the same agent \astrid{that also causes the coronal EUV wave},
we investigated whether this correlation is also present for the other BTL flares observed by the LAT. During the 24$^{\rm th}$ solar cycle three BTL flares were detected by the LAT~\citep{2017FermiBTL} and coronal waves were observed for all three of these flares. The BTL flare of 2014 January 6 unfortunately was only partially observed by the LAT due to the fact that the LAT was in the South Atlantic Anomaly for the first part of the flare and thus it is not possible to estimate the peak time of the flux. During the 25$^{\rm th}$ solar cycle the LAT has already observed two BTL flares both with coronal waves associated with the gamma-ray emission, on 2021 July 17 and the 2021 September 17 (paper in prep). We analyzed the EUV data from STEREO or SDO (depending on the angle of separation with the Earth) for these additional flares and summarize the results below.

The 2021 September 17 flare originated from an AR at S30E100\footnote{As reported by Nitta on the Heliophysics Events Knowledgebase} and was accompanied by a CME with linear speed of 
\astrid{about 1640  km s$^{-1}$.} STEREO clearly observed the coronal wave detected during this flare and we estimate the   onset time on the visible disk to be at around 04:14 UT and the derivative of the time profile peaks at 04:17 UT coinciding with peak of the LAT flux at 04:17 $\pm$ 0.5 minutes. The estimated speed of the wave following the same procedure described in Section~\ref{sec:stix_analysis} based on STEREO data is found to be 344$\pm$30 km s$^{-1}$.
 
The 2014 September 1 flare originated from an AR at N14E126, had a CME with a linear speed of 2000 km s$^{-1}$ and was associated with an X2.4 GOES class flare. The coronal wave associated with this flare was clearly seen with AIA and the onset is found to be 11:08 UT which is roughly 6 minutes later than the onset of the LAT emission. Nonetheless, we find that the derivative of the coronal wave time profile peaks at 11:14 UT within 3 minutes of the peak of the LAT flux at 11:11 UT $\pm$ 1 minute. We estimated the speed of the coronal wave associated with this event and found it to be 479$\pm$89 km s$^{-1}$.
The coronal wave of this flare was extensively studied by \cite{Grechnev_2018} and they report that there were two waves and the second one had an onset at 11:02 with an uncertainty of 1 minute, however we were able to identify only a single wave in our analysis. Evidence of particle acceleration in temporal coincidence with the coronal wave was reported by \cite{2017A&A...608A.137C}. It is important to note that the gamma-ray emission from this flare was found to have a duration of almost two hours. However, the time profile of the gamma-ray emission has a one hour gap due to the satellite night, making it very difficult to perform an in-depth study of the temporal behavior of the flux over this extended time interval. We therefore can only conclude that the coupling with the coronal wave can explain the first 18 minutes of emission and that the remaining duration requires an additional mechanism, such as the scenarios discussed in several works~\citep{Plotnikov_2017,Jin2018,2020SoPh..295...18G,wu2021}. 

The BTL flare of 2013 October 10 also originated from an AR 20$^{\circ}$ behind the visible limb, was associated with a CME with a linear speed of 1200 km s$^{-1}$ and the estimated GOES class of the X-ray flare was M4.9. STEREO detected this flare as a disk event and thus it was necessary to study the propagation of the coronal wave using SDO-AIA 193 \AA\ data. From the AIA running difference we find that the coronal wave first appears over the visible limb at approximately 07:13 UTC and the time derivative of the wave peaks at 07:16 UTC. The LAT flux peaks at 07:19 $\pm$ 1 minute. The speed of the wave is found to be 534$\pm$105 km s$^{-1}$.

We list the EUV derivative peak times and the time in which the LAT flux peaked for all four flares considered in this work in Table~\ref{tab:otherbtl_results}. The \Fermi-LAT light curves together with the coronal wave time profiles and their time derivatives for the additional three BTL flares are shown in the Appendix.
\begin{table}[h!]
    \centering
    \begin{tabular}{c|c|c}
    Flare & LAT peak flux time & EUV wave peak time  \\
             & (UT) & (UT) \\
              \hline
    \hline
    FLSF~2013-10-11 &07:19 $\pm$ 1 minute  & 07:16\\
    FLSF~2014-09-01 &11:11 $\pm$ 1 minute & 11:14\\
    FLSF~2021-07-17 &05:23 $\pm$ 1 minute & 05:21 \\
    FLSF~2021-09-17 &04:17 $\pm$0.5 minute & 04:17\\
    \end{tabular}
    \caption{EUV time derivative peak time and LAT peak flux time for the four BTL flares considered in this work. The detector used to measure the EUV wave for each of the flares are discussed in section~\ref{sec:relation}.}
    \label{tab:otherbtl_results}
\end{table}

\section{Discussion}\label{sec:discussion}

In this work we have presented the observations of the most distant BTL gamma-ray flare ever detected and have shown how the onset on the visible disk of the associated EUV coronal wave coincides with the onset of the LAT flux. Furthermore, we find compelling evidence of a coupling between the EUV coronal wave and the ion acceleration occuring during the gamma-ray flare.
While the wave speed for this event  was more consistent with a linear wave, the associated type~II radio burst proves that the disturbance was shocked at a larger height which cannot be directly seen in EUV.  We also find that the time derivative of the coronal wave profile peaks in coincidence with the peak of the $>$100~MeV flux within uncertainties for all of the BTL flares observed by the LAT. In most cases, the estimated speeds of the coronal waves associated with the LAT flares fall within the range of the population that can be explained in terms of the wave/shock model, as reported by \cite{2011A&A...532A.151W}.

Similar to what happens between hard X-ray (HXR) and soft X-ray (SXR), where the energetic electrons responsible for HXR by collisional bremsstrahlung are the main source of heating and mass supply of the SXR-emitting hot coronal plasma \citep[the so called Empirical Neupert effect,][]{1968ApJ...153L..59N,NeupertEmpirical}, the correlation between the peak of the time derivative of the coronal EUV wave and the peak of the gamma-ray flux suggests that the disturbance driving the wave is also where the ions are being accelerated and that we are probing the information on the rate of ions precipitating to the photosphere to produce gamma-rays. Combining this information with the observations of the type II burst, we can conclude that we are observing the effects of the same disturbance but at different heights and that all of these emissions share the common origin of CME-shock acceleration. 
Furthermore, there are hints in the 2021 July 17 LAT flux of a double peaked feature that also appear to be present in the derivative of the STEREO-A EUVI intensity profile that provides additional support of the connection between these two independent quantities. Unfortunately the limited LAT statistics from this flare do not allow us to reject the simple single peaked model in favor of the more complex double peak model. 
Nonetheless, the fact that this correlation is present in the data of 3 other BTL flares (as reported in Section~\ref{sec:relation}) suggests that CME-shock acceleration could also explain the gamma-ray emission in those events.

The duration of the \Fermi-LAT BTL flares are all between 10--30 minutes (with exception of the 2014 September 1 flare that shows a second longer emission phase) and their flux profiles all show a rise and fall behavior with time suggesting that we are not observing the prompt or delayed phase but a distinct phase of the gamma-ray flare. These characteristics are very similar to those of the so-called \emph{short-delayed} flares reported in the First \Fermi-LAT Solar Flare Catalog~\citep{flarecatalog_2021} where gamma-ray emission of weaker intensity with respect to the prompt emission is observed to occur between the prompt and delayed phase of the flare. The duration of these \emph{short-delayed} flares (that make up roughly 22\% of the total sample) are also of the same order as the BTL flares. The shock acceleration might very well be the source of particle acceleration for this intermediate phase of the gamma-ray flares even for on-disk flares~\cite{Kouloumvakos_2020}. Observational support for a potentially additional acceleration mechanism at work during this intermediate phase was reported by \cite{2018ApJ...865L...7O} for the 2017 September 10 flare but unfortunately only part of the \emph{short-delayed} phase was observed. Confirming this additional hypothesis could be challenging because a clear signature of the coronal wave is more easily obtained from BTL flares because the  brighter activity tied to the flare acceleration at the AR is occulted.

\begin{acknowledgments}
The \textit{Fermi} LAT Collaboration acknowledges generous ongoing support
from a number of agencies and institutes that have supported both the
development and the operation of the LAT as well as scientific data analysis.
These include the National Aeronautics and Space Administration and the
Department of Energy in the United States, the Commissariat \`a l'Energie Atomique
and the Centre National de la Recherche Scientifique / Institut National de Physique
Nucl\'eaire et de Physique des Particules in France, the Agenzia Spaziale Italiana
and the Istituto Nazionale di Fisica Nucleare in Italy, the Ministry of Education,
Culture, Sports, Science and Technology (MEXT), High Energy Accelerator Research
Organization (KEK) and Japan Aerospace Exploration Agency (JAXA) in Japan, and
the K.~A.~Wallenberg Foundation, the Swedish Research Council and the
Swedish National Space Board in Sweden.
Additional support for science analysis during the operations phase is gratefully
acknowledged from the Istituto Nazionale di Astrofisica in Italy and the Centre
National d'\'Etudes Spatiales in France. This work performed in part under DOE
Contract DE-AC02-76SF00515.

AMV acknowledges the Austrian Science Fund FWF: I4555. We thank the Gauribidanur team for sharing the GLOSS data. The solar radio facilities in the Gauribidanur observatory are operated by the Indian Institute of Astrophysics.The authors acknowledge the technical support provided by Giovanna Senatore. NO acknowledges the financial support for travel provided by the University of Pisa. This work was supported by the EU Horizon 2020 Research and Innovation Program under the Marie Sklodowska-Curie Grant Agreement 734303.
\textit{Solar Orbiter} is a space mission of international collaboration between ESA and NASA, operated by ESA. The STIX instrument is an international collaboration between Switzerland, Poland, France, Czech Republic, Germany, Austria, Ireland, and Italy.

\end{acknowledgments}

%

\vspace{5mm}
\facilities{Fermi (LAT), Solar Orbiter (STIX), Solar Dynamic Observatory (AIA), STEREO-Ahead (EUVI)}, Gauribidanur Observatory (GLOSS)
\software{
3ML \citep[\url{https://threeml.readthedocs.io/en/stable/}][]{3ML},
fermitools (\url{https://fermi.gsfc.nasa.gov/ssc/data/analysis/software/}),
astropy \citep{astropy},
SunPy (\url{https://github.com/sunpy/sunpy}),
SolarSoftWare (\url{https://www.lmsal.com/solarsoft/})}




\appendix
\label{sec:appendix}
In this appendix we report the light curves of the additional BTL flares observed by \Fermi-LAT 2013-10-11, 2014-09-01 and 2021-09-17. The multiwavelength observational overview for each of these flares are discussed in Section~\ref{sec:otherbtls}.

\begin{figure*}
\begin{center}
\includegraphics[width=0.75\textwidth]{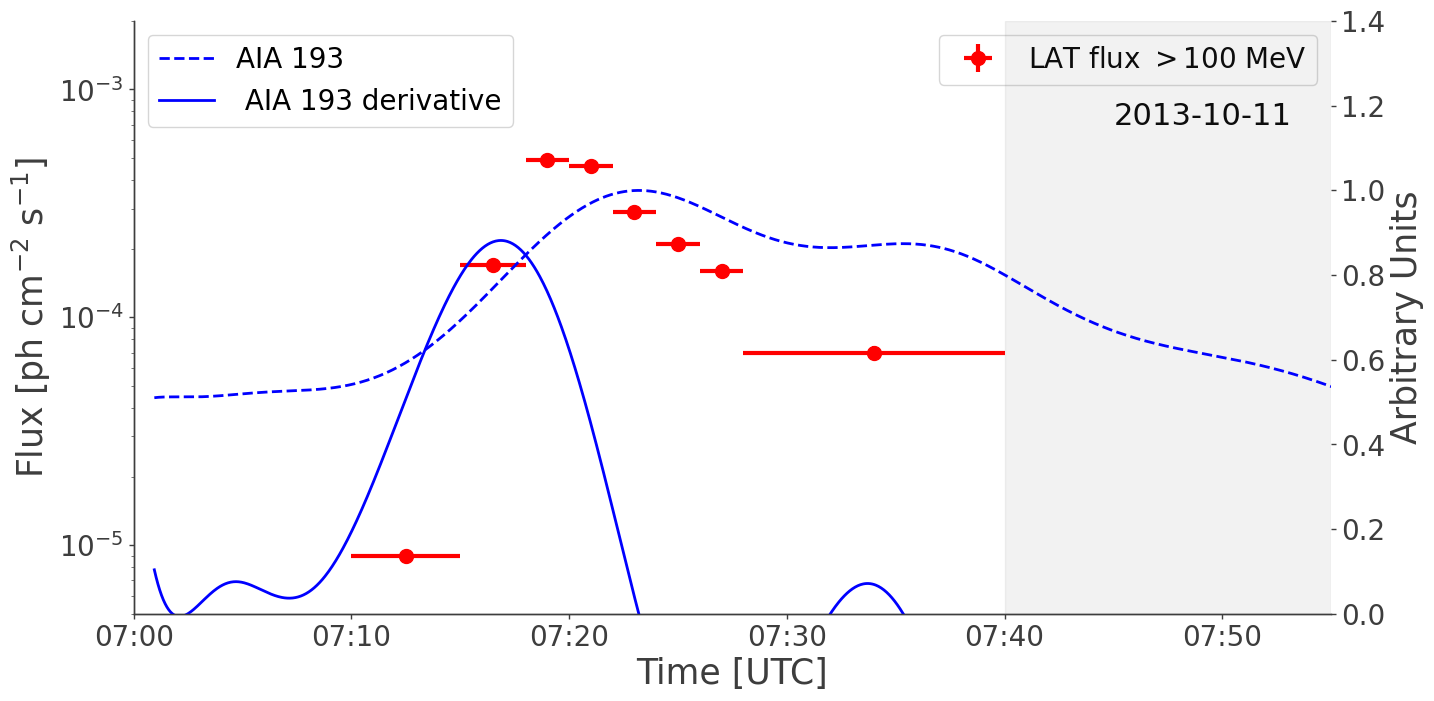}
\includegraphics[width=0.75\textwidth]{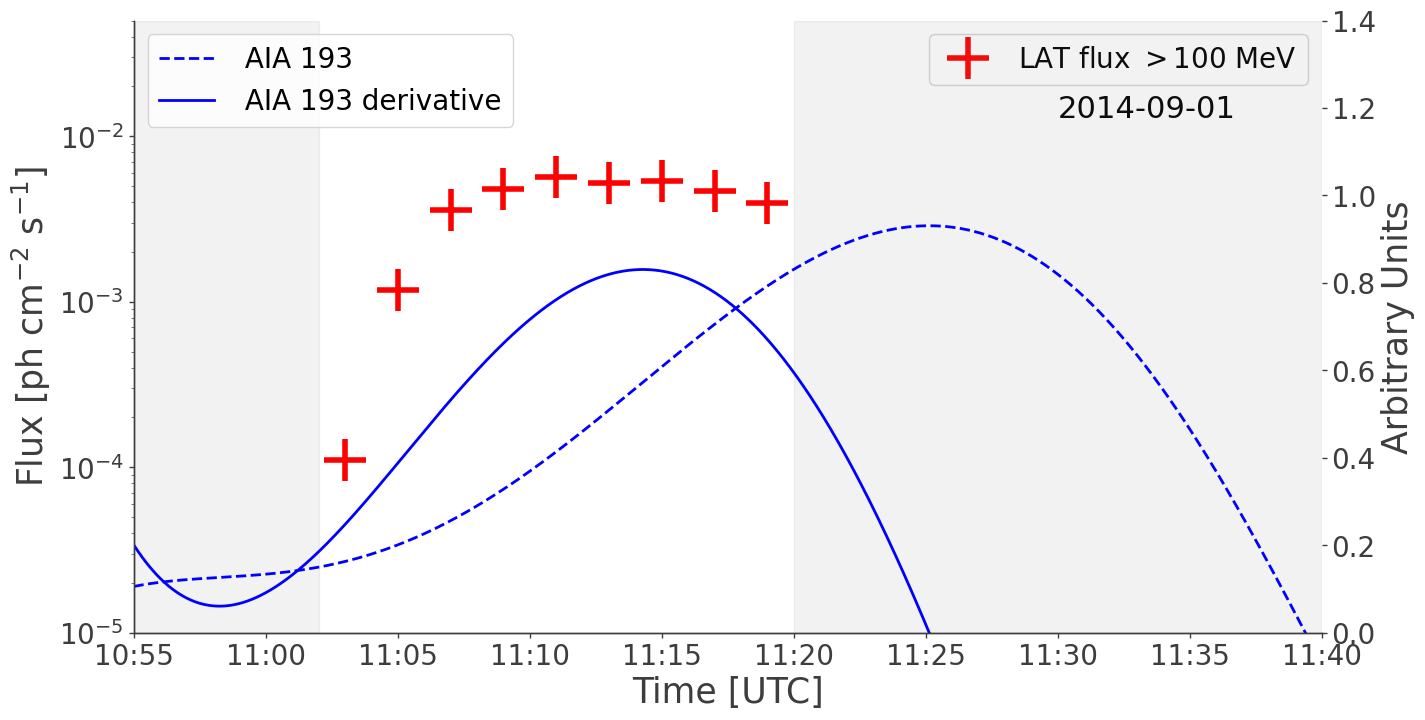}
\includegraphics[width=0.75\textwidth]{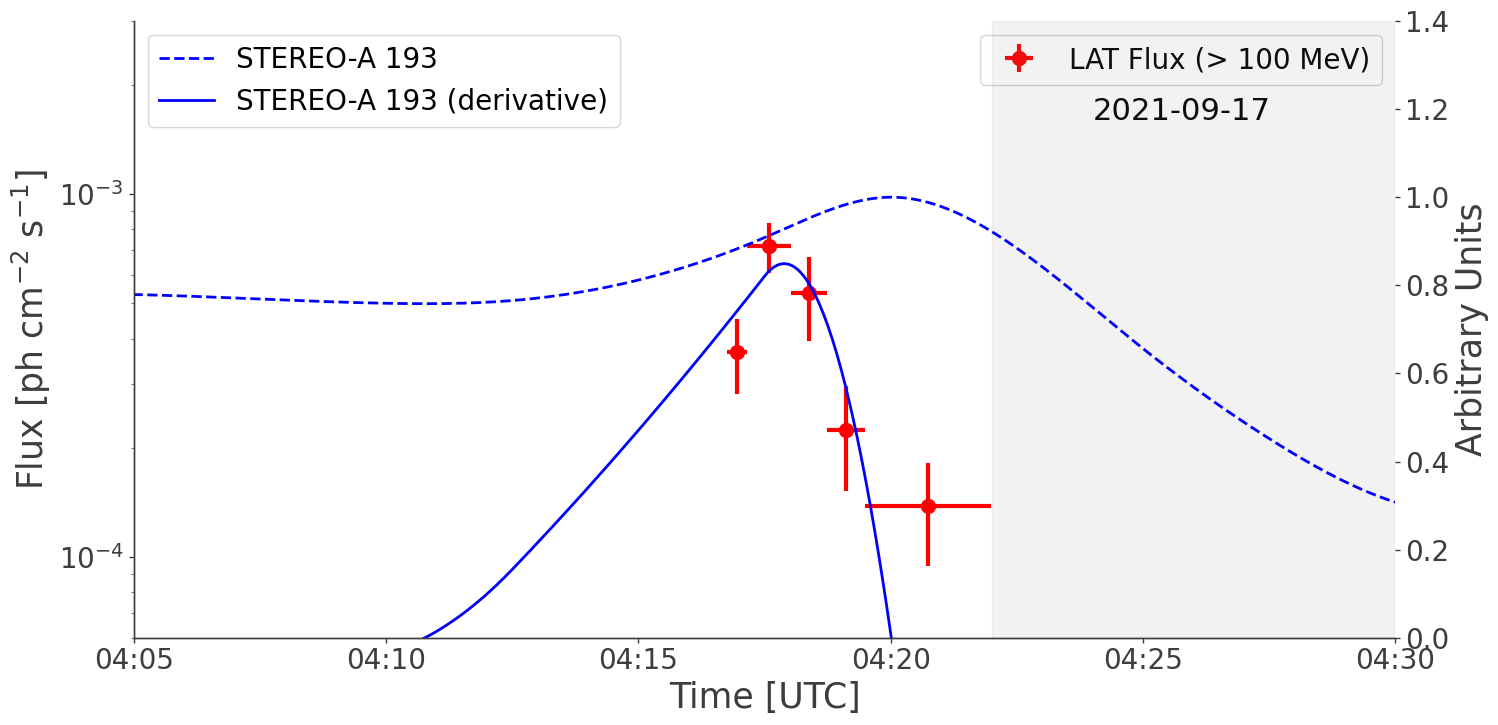}
\caption{Multiwavelength light curves for the flares of 2013-10-11 (top),  2014-09-01 (middle) and 2021-09-17 (bottom). \Fermi-LAT $>$100 MeV flux points are shown in red, coronal wave time profile in 193$\angstrom$ as measured by STEREO-A or AIA (annotated in the panel) is represented by the dashed blue lines and the time derivative of the intensity enhancement is represented by the solid blue lines. The times in which the LAT flux and the derivative of the intensity enhancement peak are reported in Table~\ref{tab:otherbtl_results}. The flux points for the flares of 2013-10-11 and 2014-09-01 are from~\cite{2017FermiBTL} while the flux points for the flare of 2021-09-17 were obtained following the same analysis procedure described in Section~\ref{sec:lat_analysis}. \journalcomments{The gray bands indicate when the Sun was out of the \Fermi-LAT FoV.}}
\label{fig:otherflares}
\end{center}
\end{figure*}

\bibliography{SOL210717}{}
\bibliographystyle{aasjournal}



\end{document}